\documentclass[pra,twocolumn,superscriptaddress,floatfix,showpacs,10pt,letterpaper]{revtex4}
\usepackage{graphicx} \usepackage{subfigure} \usepackage{amssymb}
\usepackage{verbatim} \usepackage{amsmath} \usepackage{amscd}
\usepackage{amssymb} %% Load more math symbols
\usepackage{epsfig}

\DeclareMathOperator{\tr}{tr} 
\DeclareMathOperator{\supp}{supp} 
\DeclareMathOperator{\wt}{wt} 
\DeclareMathOperator{\aut}{aut}

\begin{document}

\title{Subsystem stabilizer codes cannot have a universal set \\ of transversal gates for even one encoded qudit}

\author{Xie Chen}
\affiliation{Department of Physics, Massachusetts Institute of
Technology, Cambridge, MA 02139, USA}

\author{Hyeyoun Chung}
\affiliation{Department of Electrical Engineering, Massachusetts
Institute of Technology, Cambridge, MA 02139, USA} 

\author{Andrew W. Cross}
\affiliation{Department of Electrical Engineering, Massachusetts
Institute of Technology, Cambridge, MA 02139, USA} 
\affiliation{IBM
Research Division, T. J. Watson Research Center, P. O. Box 218,
Yorktown Heights, NY 10598}

\author{Bei Zeng}
\affiliation{Department of Physics, Massachusetts Institute of
Technology, Cambridge, MA 02139, USA}

\author{Isaac L. Chuang}
\affiliation{Department of Physics, Massachusetts Institute of
Technology, Cambridge, MA 02139, USA} 
\affiliation{Department of
Electrical Engineering, Massachusetts Institute of Technology,
Cambridge, MA 02139, USA}

\date{\today}

\begin{abstract}  
A long-standing open problem in fault-tolerant quantum computation has been to find a universal set of transversal gates. As three of us proved in arXiv: 0706.1382, such a set does not exist for binary stabilizer codes. Here we generalize our work to show that for subsystem stabilizer codes in $d$ dimensional Hilbert space, such a universal set of transversal gates cannot exist for even one encoded qudit, for any dimension $d$, prime or nonprime. This result strongly supports the idea that other primitives, such as quantum teleportation, are necessary for universal fault-tolerant quantum computation, and may be an important factor for fault tolerance noise thresholds. 
\end{abstract}
 \maketitle %\narrowtext

%%%%%%%%%%%%%%%%%%%%%%%%%%%%%%%%%%%%%%%%%%%%%%%%%%%%%%%%%%%%%%%%%%%%%%%%%%%%%

\section{Introduction}

All quantum systems are vulnerable to noise, which can arise from various sources such as uncontrolled interactions of the system with the environment, or from imperfections in the implementation of quantum logical operations. Moreover, noise can propagate through a quantum circuit, affecting qudits throughout the computational system. Thus, if quantum computation is to be implemented on a large scale, it is essential to explore methods for protecting quantum information against noise, and for preventing the spread of errors through a quantum system.

The theory of quantum error-correcting codes, coupled with fault-tolerant quantum computation, offers the hope of resolving both of these problems, and has therefore greatly improved the long-term prospects for quantum computing technology \cite{Nielsen,Preskill}. Quantum error-correcting codes encode quantum information in a form that is more resistant to noise. In principle, fault-tolerant quantum computation then makes it possible to carry out arbitrarily long quantum computations reliably, provided that the average probability of error per gate is less than a certain critical value known as the accuracy threshold \cite{Gottesman}.

The precise implementation of fault-tolerant quantum computation depends on the quantum error-correcting code (QECC) used. Once a QECC has been chosen, each qubit in the original circuit is replaced with an encoded block of qubits. Protocols are then specified for performing fault-tolerant operations on the code, i.e. protocols for each type of logic gates. A fault-tolerant protocol is designed so that if only one component of a procedure fails, then the failure causes at most a correctable number of errors in each encoded block of qudits output from the procedure.

Many studies concentrate only on the case of binary QECCs in a $d=2$ dimensional Hilbert space, as generalizations of proofs are often non-trivial when $d>2$ is nonprime. However, as both qubit and qudit systems occur in the natural world, there is no reason to assume that a theoretical result should hold solely for $2$-dimensional systems. If an important negative result were to hold only in the case when $d=2$, then this would suggest that a lot of effort should be directed towards building qudit systems, as the case when $d>2$ would be fundamentally different from the case $d=2$ (for instance, quantum error-correcting codes of certain parameters only exist in the case of $d>2$ \cite{Markus}). Therefore, it is important to consider the case of higher dimensional systems, and in our work we consider the case of QECCs for arbitrary $d$, both prime and nonprime.

One way of implementing fault-tolerant quantum operations is to use transversal gates \cite{Gottesman}. A transversal gate has a particularly simple form: it is a tensor product of unitaries that each act on only one qudit per encoded block \cite{Shor}. Thus, transversal gates are naturally designed to limit the propagation of noise, as an error occuring on the $k$th qudit in a block can only ever propagate to the $k$th qudit of other blocks of the code, no matter what other sequences of gates we perform before retrieving the encoded information.

As transversal gates offer significant advantages in constructing fault-tolerant quantum circuits, it is highly desirable to know exactly which gates can be performed transversally on a given QECC. In the case of certain codes, such as the $7$-qudit Steane code for $d=2$, a number of different gates can be performed transversally: in particular, any gate from the Clifford group can be implemented as a transversal gate. It would be wonderful to find a QECC such that universal quantum computation can be achieved entirely through transversal operations on the code. Unfortunately, it is widely believed in the quantum information science community that no such code exists \cite{Gottesman}. 

A proof of this belief is of fundamental importance in the fault-tolerant design of quantum circuits and the estimation of the accuracy threshold, as such a proof would provide valuable information about the fundamental resources needed for quantum computation. If there is no QECC such that a universal set of gates can be performed transversally on the code, then transversal gates are not the ultimate primitives for fault-tolerant universal quantum computation: they must be supplemented with more complicated techniques, such as quantum teleportation \cite{GotChuang,KnillInject} or state distillation \cite{BravyiDistill}.

Several difficulties must be overcome in order to prove that transversality is insufficient for universality. Even though the gates that can be implemented transversally on a given QECC depend on the code itself, the result must hold for all error-correcting codes. Furthermore, the logical operation of the gate on the encoded information must be determined from the physical operation of a transversal gate on the qudits of a quantum system. Finally, the important step of generalizing this result for qudits in a Hilbert space of arbitrary dimension $d$ is not necessarily straightforward, particularly if $d$ is nonprime.

In this paper, we approach the problem of proving that stabilizer codes cannot have a universal set of transversal gates. Recently, three of us proved in \cite{Bei} that a universal set of transversal gates does not exist for binary stabilizer codes. Here we generalize our earlier result by showing the following Main Theorem.

{\bf Main Theorem}: {\it For subsystem stabilizer codes in $d$ dimensional Hilbert space, a universal set of transversal gates cannot exist for even one encoded qudit, for any dimension $d$, prime or nonprime.} 

Given that stabilizer codes form the most important and well-developed class of quantum error-correcting codes, the situation considered in our proof is very general. We also provide an alternative insight into the problem by introducing a different proof technique from the one given in \cite{Bei}, which uses an idea in a recent work by Daniel Gross and Maarten Van den Nest \cite{Gross}. This technique is more transparent and accessible than the approach taken in \cite{Bei}, and provides more intuition for the final result.

Our proof has two main stages. Firstly, given a transversal gate $U=\otimes_{j=1}^n{U}_j$, we show that there must be some restrictions on the physical operations $U_j$ for every $j$. We derive these restrictions by studying the subcodes of a stabilizer code. This idea dates back to work carried out by Rains \cite{RainsAut}, who showed that any transversal gate on a given stabilizer code must keep some subcodes invariant. This fact allows us to place strong conditions on the structure of the transversal gate. After describing these restrictions on the operations $U_j$, we then show that there must be corresponding restrictions on the logical operation $U$ that prevent universality.

Our paper is organized as follows: we begin by presenting some background information on the generalized Pauli group, stabilizer codes, and transversal gates in Sec. \ref{Sec-Preliminaries}. We then identify the restrictions on the structure of transversal gates in Sec. \ref{Sec-Binary} (the binary case $d=2$) and Sec. \ref{Sec-NonBinary} (the nonbinary case $d > 2$). In Sec. \ref{Sec-Logical} we analyze the effect of the restrictions on the transversal operations on the logical gate. We conclude in Sec. \ref{Sec-Conclusion} with a discussion of open problems, in particular the effect of coordinate permutation on the possibility of achieving universality using transversal gates.

\section{Preliminaries}\label{Sec-Preliminaries}

\subsection{The generalized Pauli group}

In this section we review the mathematics for the generalized Pauli group ${\mathcal P}^d$,
which will be the main mathematical tool for describing the
qudit stabilizer codes. The generalized Pauli group is generated by two elements $X,Z$ with the commutation relation \cite{Sun,Bar1,San,Dab,Got,Pat,Kni}
\begin{equation}
ZX=qXZ, \label{qpc}
\end{equation}
where $q$ is a complex number. Mathematically, we can prove
that the associated group generated by $Z$, $X$  possesses a
$d-$dimensional irreducible representation only for $q^{d}=1$
\cite{Sun}. In this article, we take $q\equiv q_d\equiv e^{i \frac
{2\pi}{d}}$. This special case was first introduced by Weyl
\cite{Wey}, and its completeness was first proved by Schwinger
\cite{Sch}. Obviously, when $d=q=1$, the generators $X$ and $Z$ can be
regarded as the ordinary coordinates of ${\mathbb R}^{2}$ plane. When
$d=2,\,q=-1$, the generators $X$ and $Z$ can be identified with the Pauli
matrices $\sigma _{x}$ and $\sigma _{z}$, and the generalized Pauli group ${\mathcal P}^2$ is the familiar $1$-qubit Pauli group, also denoted by ${\mathcal P}$. 

Choosing a basis $|k\rangle_{k=0}^{d-1}$, we have

\begin{equation}
Z|k\rangle=q_d^k|k\rangle,
\end{equation}
where $|k\rangle =X^{\dagger k}|0\rangle$. This also implies
\begin{equation}
X|k\rangle =|k+1\rangle. \label{xiz}
\end{equation}
In the $Z$-diagonal representation, the matrices of $X$ and $Z$
are:
\begin{equation}
Z=\left[
\begin{array}{cccccc}
1 & 0 & 0 & \cdots & 0&0 \\
0 & q_d & 0 & \cdots & 0&0 \\
\vdots & \vdots & \vdots& \ddots  & \vdots & \vdots\\
0 & 0 & 0 & \cdots & q_d^{d-2}&0 \\
0 & 0 & 0 & \cdots & 0& q_d^{d-1}
\end{array}
\right],
\end{equation}
\begin{equation}
X=\left[
\begin{array}{cccccc}
0 & 0 & 0 & \cdots &0& 1\\
1 & 0 & 0 & \cdots &0& 0 \\
0 & 1 & 0 & \cdots & 0&0 \\
\vdots & \vdots & \vdots& \ddots  & \vdots & \vdots\\
0 & 0 & 0 & \cdots &1& 0 
\end{array}
\right].
\end{equation}

All the elements of the generalized Pauli group are given by
\begin{equation}
\{q_d^iZ^{j}X^{k} \ | \ i, j, k\in {\mathbb Z}_{d}\}
\end{equation}

Define a {\bf basis set} $\mathcal{B}_d$ of ${\mathcal P}^d$ by
\begin{equation}
\mathcal{B}_d=\{Z^{j}X^{k} \ | \ j, k\in {\mathbb Z}_{d}\},\label{oba}
\end{equation}
and call the elements in $\mathcal{B}_d$ the {\bf basis elements} of ${\mathcal P}^d$,
then the general commutation relations for any two basis elements are
\begin{equation}
Z^{j}X^{k}=q_d^{jk} X^{k}Z^{j}.  \label{unibase}
\end{equation}

In addition, we can replace the generators $Z$ and $X$ with two
other elements in the basis. First, let $(m,n)$ denote the greatest
common factor of integers $m$ and $n$. Then if
$(m_1,n_1)=1$ for $m_1,n_1\in Z_d$, we can define
\begin{equation}
\bar{X}=q_d^{-\frac{d-1}{2}m_1n_1}Z^{m_1}X^{n_1},
\end{equation}
where the factor before $Z^m X^n$ is chosen so that $\bar{X}$ has the same
eigenvalues as $X$. To maintain Eq. (\ref{qpc}),  we define
\begin{equation}
\bar{Z}=q_d^{-\frac{d-1}{2}m_2n_2}Z^{m_2}X^{n_2},
\end{equation}
where $(m_2,n_2)=1$ for $m_2,n_2\in Z_d$, and $m_1n_2-m_2n_1=1$.
From another viewpoint, $\bar{X}$ and $\bar{Z}$ define a unitary
transformation $U$ such that
\begin{equation}
\bar{X}=UXU^{\dagger}, \quad \bar{Z}=UZU^{\dagger}.
\end{equation}
By the above definition, it is easy to check that the set of all such unitary transformations $U$ forms a group, which is known as the Clifford group.

Finally, we define the $n$-qudit Pauli group. The familiar $n$-qubit Pauli group ${\mathcal P}_n$ consists of all local operators of the form $R = \alpha_R R^{(1)}\dots R^{(n)}$, where $\alpha_R \in \{\pm 1, \pm i\}$ is an overall phase factor and $R^{(i)}$ is either the identity $I$ or one of the Pauli matrices $\sigma_x, \sigma_y$, or $\sigma_z$. We can define the analogous $n$-qudit Pauli group ${\mathcal P}_n^d$ as the set of all local operators of the form $R = \alpha_R R^{(1)}\dots R^{(n)}$, where $\alpha_R = q_d^k$ for some $k \in {\mathbb Z}_d$ is an overall phase factor and $R^{(i)}$ is an element of the generalized Pauli group ${\mathcal P}^d$.

\subsection{Stabilizer codes, transversal gates, and encoded universality}\label{Sec-Prelim2}

In this section we introduce definitions and notation for studying stabilizer codes and transversal gates. Let $Q$ denote an $[[n,k,\delta]]$ binary stabilizer code with stabilizer $S$ \cite{CRSS,Gottesman}. The orthogonal projector onto $Q$ is denoted by $P_Q$ and is given by
\begin{equation}
P_Q = \frac{1}{2^n}\sum_{R\in S}R.
\end{equation}

The code may or may not be a subsystem code \cite{Poulin}. If the code is a subsystem code, there are $k'\geq 0$ 
additional logical qubits, $S$ is generated by $n-k-k'$ independent generators, and the corresponding
subspace code is an $[[n,k+k',\delta']]$ code with $\delta'\leq \delta$. The $k'$ additional logical qubits are known as the {\bf gauge qubits}, and the original $k$ logical qudits are known as the {\bf protected qubits}.

A qudit stabilizer code $Q_d$ is then the vector space stabilized by a subgroup $S$ of the generalized Pauli group, such that $q_d^l I \notin S$ for $l \neq 0$. An $[[n,k,\delta]]$ stabilizer code encodes $k$ logical \textit{qudits} into $n$ physical \textit{qudits} and can correct up to $\lfloor\frac{\delta-1}{2}\rfloor$ independent single qudit errors.

\begin{figure}%[htb!]
\centering%
\includegraphics[width=2in]{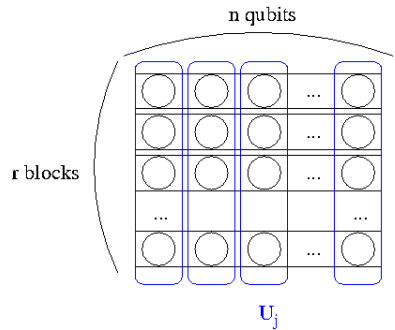}
\caption{Illustration of a transversal gate on $r$ blocks of $n$ qubits each. The blocks are represented by
a collection of circles (qubits), grouped into boxes of $n$. The $r$ blocks undergo a transversal gate
whose unitaries $U_j$ act on qubits in the [blue] boxes with rounded edges.}
\label{fig:transversal}
\end{figure}

Suppose there are initially $r$ blocks of $k$ qudits in a $d$-dimensional Hilbert space and we encode each block of $k$ qudits
into $Q$. In order to define a transversal gate acting on these $r$ blocks, we must first define the {\bf local unitary group}. For the single block case, the local unitary group is $G=U(1)\times SU(d)^n$. Each state $P_Q$ has a {\bf stabilizer subgroup} $I_{Q} \subset G$ consisting of elements $g \in G$ that leave $P_Q$ fixed under the action $gP_Q g^{-1}$ \cite{Lyons}. For the multiblock case with $r$ blocks, the local unitary group is $G_r=U(1)\times SU(d^r)^n$. Each state $P_Q^{\otimes r}$ has a stabilizer subgroup $I_{Q}^r \subset G_r$ consisting of elements $g \in G_r$ that leave $P_Q^{\otimes r}$ fixed under the action $gP_Q^{\otimes r}g^{-1}$. The subgroup $I_{Q}^r$ is known as the local unitary group of $P_{Q}^{\otimes r}$. A \textbf{transversal gate} acting on the $r$ blocks is an $nr$ qudit unitary $U$ that is an element of the local unitary group $I_Q^r$ of $P_Q^{\otimes r}$. The gate factors into an $n$-fold tensor product $U = \otimes_{j=1}^n U_j$ of $r$ qudit unitaries $U_j$. Each $U_j$ acts on the $j$th qudit of the $r$ blocks. See FIG.~\ref{fig:transversal} for an illustration of a transversal gate applied to $r$ encoded blocks of $n$ qubits (the case $d=2$) each.

\subsection{Problem and Strategy}\label{Sec-Prelim3}

We would like to know if the transversal gates are an encoded quantum computationally \textbf{universal set} for 
at least one of the encoded qudits. If so, then this means that it is possible to approximate any single qudit 
logical gate on one of the $k$ encoded qudits (we don't care which one) to any accuracy using only transversal 
gates. Since the transversal gates form a group, we can formally state this as follows: Given any encoded single qudit unitary gate 
$V$ on a fixed encoded qudit and an accuracy $\epsilon>0$, there is a transversal $r$-block gate $U_\epsilon$ such 
that $||U_\epsilon P_Q^{\otimes r} - VP_Q^{\otimes r}||<\epsilon$. We will assume that this is true and derive a contradiction, which implies that the transversal gates are not computationally universal.

Our general strategy is to show that the condition of transversality places restrictions on the form of each $U_j$ in the tensor product expansion $U = \otimes_{j=1}^n U_j$ of a transversal gate. In Sec. \ref{Sec-Binary} and \ref{Sec-NonBinary} we derive the exact forms of these restrictions for the cases when $d=2$ and $d>2$, respectively. In Sec. \ref{Sec-Logical} we use these results to show that the restrictions on the $U_j$ place enough constraints on the logical operation $U$ to prevent universality.

\section{The structure of stabilizer subgroups of stabilizer codes: binary case}\label{Sec-Binary}

In this section we show that a transversal gate acting on $r$ blocks of $n$ qubits encoded using a stabilizer code $Q$ has a severely restricted form. We first introduce some definitions that allow us to formally state the restrictions on transversal gates. An $n$-qubit unitary operation is said to be {\bf semi-Clifford} if it sends at least one maximal abelian subgroup of the $n$-qubit Pauli group ${\mathcal P}_n$ to another maximal abelian subgroup of ${\mathcal P}_n$ under conjugation. If $T$ is a semi-Clifford operation, then there exist Clifford operations $L_1,L_2$ such that $L_1TL_2$ is diagonal \cite{ZCC}.

An $n$-qubit unitary operation is said to be {\bf generalized semi-Clifford} if it sends the span of one maximal abelian subgroup of ${\mathcal P}_n$ to the span of another maximal abelian subgroup of ${\mathcal P}_n$ under conjugation. If $T$ is a generalized semi-Clifford operation, then there exist Clifford operations $L_1,L_2$, and a classical permutation operator $P$ such that $PL_1TL_2$ is diagonal \cite{ZCC}.

Our main task in this section is to prove the following theorem.

{\bf Theorem 1:} Given an $n$-qubit stabilizer code $Q$ free of Bell pairs and trivially encoded qubits, let $U = \otimes_{j=1}^n U_j$ be an element of $I_{Q}^r$. Let $[n]$ denote the 
set of positive integers from $1$ to $n$. Then for each $j\in [n]$, $U_j$ is an $r$-qubit generalized semi-Clifford operation.

This theorem places severe restrictions on the physical form of a transversal gate $U$. In Section \ref{Sec-Logical}, we will show that these restrictions place corresponding constraints on the logical gate $U$, thereby making it impossible to achieve universality using only transversal gates.

Proving this theorem is not trivial, as we must draw conclusions about each factor $U_j$ of the transversal gate $U$, given only information about the action of $U$ on the entire codespace. We will prove the theorem by studying codes that are stabilized by subgroups of $S$. Such a code is known as a stabilizer {\bf subcode}. We can show that a transversal gate preserves certain stabilizer subcodes. This requirement allows us to place restrictions on the form of transversal gates by studying subcodes of a special form. The following important lemma will be useful in studying the action of transversal gates on stabilizer subcodes.

{\bf Lemma 1:} Let $\omega\subseteq [n]$ be a nonempty subset of coordinates, and let $\bar{\omega}$ denote the set $[n]\setminus \omega$. Given a transversal gate $U = \otimes_{i=1}^n U_i$, let $U_\omega\equiv\bigotimes_{i\in\omega} U_i$. We can then write
\begin{align}\label{Eq-Lemma1}
\tr_{\bar{\omega}}\left[ UP_Q^{\otimes r}U^\dag \right] &=\rho_\omega^{\otimes r},
\end{align}
where $\rho_\omega$ is defined as
$\tr_{\bar{\omega}}P_Q$.

To prove the lemma, note that since a transversal 
gate $U$ is an encoded gate, we can write
\begin{align}
\tr_{\bar{\omega}}\left[ UP_Q^{\otimes r}U^\dag \right] &= U_{\omega}\tr_{\bar{\omega}}\left[ P_Q^{\otimes r}\right]U_{\omega}^\dag\nonumber\\
&= U_\omega \rho_\omega^{\otimes r} U_\omega^\dag
= \rho_\omega^{\otimes r},
\end{align}
which gives the necessary result.

This lemma tells us that an encoded gate also preserves the subcodes $\rho_\omega^{\otimes r}$ for any $\omega$.
This result is useful because we can turn it around -- if a gate does not preserve subcodes, then it cannot be an encoded gate.
Note that it is easy to compute the projector $\rho_\omega$ onto the subcode from the projector $P_Q$ onto the original 
code. We define the {\bf support} $\supp(R)$ of an element $R \in S$ to be the set of all $i \in [n]$ such that the $i$th coordinate $R^{(i)}$ differs from the identity. We say that an element $R \in S$ has {\bf full support} if $\supp(R) = [n]$. We then have
\begin{align}
\rho_\omega &= \tr_{\bar{\omega}} P_Q
\propto \tr_{\bar{\omega}} (\sum_{R\in S} R)\nonumber\\
&= \sum_{R\in S} \tr_{\bar{\omega}} R
= \sum_{R\in S,\, \supp(R)\subseteq\omega} R.\label{Eq-StabilizerSubcode}
\end{align}

The set $S_\omega=\{ R\in S\ |\ \supp(R)\subseteq\omega\}$ is the stabilizer of the subcode, which is a subgroup of $S$. The partial trace removes the unencoded qubits at coordinates in $\bar{\omega}$ from the subcode. 

We will prove Theorem 1 in two ways, by studying two classes of stabilizer subcodes. In Sec. \ref{Sec-MinimalSubcodesBinary} we use the so-called minimal subcodes of $S$, and in Sec. \ref{Sec-OneQubitSubcodesBinary} we use subcodes associated with single qubits. For the rest of this section we will work with an $n$-qubit stabilizer code $Q$ with corresponding stabilizer $S$ that satisfies the conditions of Theorem 1.

\subsection{Minimal subcodes and beyond}\label{Sec-MinimalSubcodesBinary}

\subsubsection{Minimal subcodes}

In order to define minimal subcodes, we must first introduce the concept of \textbf{minimal supports}. A support $\omega$ is a minimal support of $S$ if there is a nonidentity element of $S$ with support $\omega$, and there are no elements with support strictly contained in $\omega$. An element in $S$ with minimal support is called a \textbf{minimal element} \cite{RainsAut}.
The concept of a minimal support of a stabilizer state has been extremely useful in the study of local unitary
versus local Clifford equivalence of stabilizer and graph states \cite{Nest,Zeng}.
Minimal supports have also arisen in classical coding theory in the context of secret sharing schemes \cite{Ashikhmin},

Given a minimal support $\omega$, then all the nonidentity elements in $S_{\omega}$ have support $\omega$. The following lemma of Van den Nest \cite{Nest} allows us to characterize $S_{\omega}$ for a minimal $\omega$.

{\bf Lemma 2:} Let $A_\omega$ denote the number of nonidentity elements in $S_{\omega}$ with minimal support $\omega$. Then $A_\omega = 1$ or $3$.

The proof is fairly straightforward. By definition, there must be some element of $S$ with support $\omega$, so if there are no more, $A_\omega=1$. If there are two elements $M,N$ with support $\omega$, then their product $MN$ must have support $\omega$ too, as otherwise $\omega$ is not minimal. So $A_\omega$ cannot be 2, but it can be 3. Suppose there is a fourth element $M'$ with support $\omega$. There are only three nonidentity Pauli 
operators, so one of them must appear twice at some coordinate in $\omega$.  But then we can form another product whose 
support is strictly contained in $\omega$, meaning that $\omega$ is not a minimal support, so $A_\omega$ cannot be greater than 3. Notice that when $A_\omega=3$, $|\omega|$ must be even, otherwise the operators will not commute.

We can use this result to describe the subcode stabilized by $S_\omega$. By Lemma 2, $S_\omega$ has either $2$ or $4$ elements. We denote the coordinates in $\omega$ by $j\in\{1,2,\dots,|\omega|\}$, though we will understand that this notation just indexes $\omega$ -- the actual coordinate is the $j$th element of $\omega$. Computing the projector $\rho_\omega$ onto the subcode stabilized by $S_\omega$, we find that either
\begin{align}
\rho_\omega &\propto \underbrace{I\otimes\dots\otimes I}_{|\omega|\ \textrm{times}} + M_1\otimes M_2\otimes\dots\otimes M_{|\omega|} \nonumber\\
&= I^{|\omega|} + M_\omega
\end{align}
or
\begin{equation}
\rho_\omega \propto I^{|\omega|} + M_\omega + N_\omega + (MN)_\omega,
\end{equation}
where $M_\omega$ and $N_\omega$ are Pauli operators in $S$ restricted to $\omega$ whose product also
has support on $\omega$. It is helpful to realize that these operators are projectors onto 
$[[|\omega|,|\omega|-1,1]]$ and $[[|\omega|,|\omega|-2,2]]$ stabilizer codes, respectively. We can also see that there is some Clifford operation that we can apply at each coordinate in $\omega$ to transform the stabilizers of these subcodes into $\langle Z^\omega\rangle$ and $\langle X^\omega, Z^\omega\rangle$, respectively. These codes are the \textbf{minimal subcodes} associated with the minimal support $\omega$.

The extent to which a stabilizer code can be described by its minimal subcodes depends on the particular stabilizer
code. For example, the \textit{$GF(4)$-linear codes} are one family of stabilizer codes that can be described completely by their minimal subcodes \cite{RainsAut,Nest}.

\subsubsection{Transversal gates on minimal subcodes}\label{Sec-MinSubcode2}

In this section, we place restrictions on the operators $U_j$ of a transversal gate $U=\otimes_{j=1}^n U_j$ when $j$ is contained in some minimal support of $S$.

Suppose we can find minimal elements whose supports cover a subset of coordinates $m\subseteq [n]$. What can we learn about the form of a transversal gate on the coordinates in $m$ by studying its action on minimal subcodes? The following discussion is a generalization of Rains' approach \cite{RainsAut}. First, recall that Clifford gates are not universal, and if we have a transversal gate constructed from Clifford gates, then that transversal gate must be some kind of logical Clifford gate as well. The challenging behavior comes from non-Clifford gates. Therefore, we will find it 
convenient to more or less ignore Clifford gates altogether. We will move to locally Clifford equivalent stabilizer codes 
freely when studying particular minimal subcodes. Keeping this in mind, we can write the $r$ block projectors when 
$A_\omega=1$ and $A_\omega=3$. If $A_\omega=1$, then
\begin{align}\label{Eq-MinSubcodeProjector1}
\rho_\omega^{\otimes r} \propto (I^\omega + Z^\omega)^{\otimes r} & = \sum_{i\in \{0,1\}^r} (Z^\omega)^{i_1}\otimes\dots\otimes (Z^\omega)^{i_r} \nonumber\\
& = \sum_{i\in \{0,1\}^r} Z(i)^{\otimes |\omega|}
\end{align}
where $i_j$ denotes the $j$th bit of $i$, in the second expression, and $Z(i)=\otimes_{j=1}^{r}Z^{i_j}$ in the third 
expression. The $Z(i)$ are the {\bf Pauli Z operators}, and form a maximal abelian subgroup of the $r$ qubit Pauli group. We can define the Pauli X and Pauli Y operators analogously. 

\begin{figure}[htb!]
\centering
\includegraphics[width=2in]{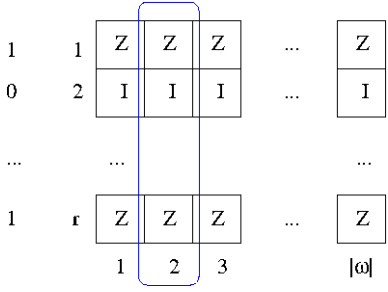}
\caption{Illustration of a single term in the expansion of $\rho_\omega^{\otimes r}$ for the case $A_\omega=1$. Each 
box is associated to a qubit in FIG.~\ref{fig:transversal}, and the value of the bit $i$ to the left of the $j$th row determines whether that row is 
$Z^{|\omega|}$ or $I^{|\omega|}$. Therefore, the Pauli $Z$ operator along each column is the same operator $Z(i)$, and it is determined 
by the bit string $i$. A factor $U_j$ of a transversal gate acts on a column (the [blue] box with rounded edges, for example).}
\label{fig:table}
\end{figure}

It may be helpful to consult FIG.~\ref{fig:table} for an illustration of one of the summands in Equation (\ref{Eq-MinSubcodeProjector1}) as it would 
look overlayed on FIG.~\ref{fig:transversal}. The third expression may be somewhat confusing because the tensor 
product ``$\otimes |\omega|$'' is over the columns of FIG.~\ref{fig:table}. We do this because the transversal gate, 
which we will apply shortly, factors into a tensor product over columns too. Similarly, if $A_\omega=3$, then
\begin{align}
\rho_\omega^{\otimes r} & \propto (I^\omega + X^\omega + Z^\omega + (-1)^{|\omega|/2}Y^\omega)^{\otimes r} \nonumber\\
& = \sum_{(a|b)\in \{0,1\}^{2r}} \left[ (-1)^{|\omega|/2}\right]^{\wt(a\cdot b)} R^\omega(a_1,b_1)\otimes\nonumber\\ &\dots\otimes R^\omega(a_r,b_r) \nonumber\\
& = \sum_{(a|b)\in \{0,1\}^{2r}} \left[ (-1)^{|\omega|/2}\right]^{\wt(a\cdot b)} R(a,b)^{\otimes |\omega|},
\end{align}
where $R(0,0)=I$, $R(0,1)=Z$, $R(1,0)=X$, and $R(1,1)=Y$, (i.e. $R(a_j,b_j)=i^{a_j\cdot b_j}X^{a_j}Z^{b_j}$) and also $R(a,b)=\otimes_{j=1}^r R(a_j,b_j)$.
Again, the tensor product in the third expression is over columns rather than rows.

One or both of the projectors we have written are left unchanged by transversal gates when the gates are restricted to a 
minimal support $\omega$, i.e. $U_\omega\rho_\omega U_\omega^\dag=\rho_\omega$. Since $U_\omega IU_\omega^\dag=I$, we can 
subtract the identity from each side from the preceding equation. Rains has shown that it is convenient to view the projectors as vectors in 
Euclidean space acted on by rotations. This association will let us show that rotations fixing these vectors have a 
special form. The $r$ qubit gate $U_j$ acts by conjugation on a nonidentity $r$ qubit Pauli matrix $R_s$ ($s$ indexes 
the $4^r-1$ nonidentity Paulis) as
\begin{equation}
U_jR_sU_j^\dag = \sum_{R_t\in {\mathcal P}_r-\{I\}} \alpha_{ts} R_t.
\end{equation}
Here ${\mathcal P}_r$ denotes the $r$ qubit Pauli group. The identity matrix does not appear on the right hand side because $U_j$ 
is unitary and $R_s$ is traceless, so the image must be traceless. The coefficients $\alpha_{ts}$ must be real because 
$R_s$ is Hermitian. Furthermore, $\sum_{R_t\in {\mathcal P}_r-\{I\}} \alpha_{ts_1}\alpha_{ts_2}=\delta_{s_1s_2}$ because $R_s$ is unitary.
So, we can represent $U_j$ by a matrix $O_j$ in $SO(4^r-1)$ whose real
entries are $\alpha_{ts}$, $s,t\in [4^r-1]$, and whose columns are orthonormal. The inverse unitary $U_j^\dag$ is represented
by the transpose $O_j^T$ and its columns are orthonormal, so both the rows and columns are orthonormal.
We can represent the nonidentity $r$ qubit Pauli matrices by the canonical 
basis vectors $\{|1\rangle,|2\rangle,\dots,|4^r-1\rangle\}$ of ${\mathbb R}^{4^r-1}$. For concreteness, we can associate the label $i$ of 
$|i\rangle$ to the binary representation $(a|b)\in\{0,1\}^{2r}$ or to the Pauli representation $i^{\wt(a\cdot b)}X(a)Z(b)$. Continuing, we can now write the subcode projectors as vectors in 
$({\mathbb R}^{4^r-1})^{\otimes |\omega|}$, using ``$\mapsto$'' to denote this mapping. For $A_\omega=1$,
\begin{equation}
\rho_\omega^{\otimes r} - I \mapsto \sum_{i=1}^{2^r-1} \underbrace{|ii\dots i\rangle}_{|\omega|\ \textrm{times}} =: w
\end{equation}
and for $A_\omega=3$,
\begin{equation}
\rho_\omega^{\otimes r} - I \mapsto \sum_{i=1}^{4^r-1} \alpha_i\underbrace{|ii\dots i\rangle}_{|\omega|\ \textrm{times}} =: v,
\end{equation}
where $\alpha_j\in \{\pm 1\}$. We can now compute
\begin{align}
ww^T = \sum_{i,j=1}^{2^r-1} |ii\dots i\rangle\langle jj\dots j|, \\
vv^T = \sum_{i,j=1}^{4^r-1}\alpha_i\alpha_j |ii\dots i\rangle\langle jj\dots j|.
\end{align}
Following Rains, consider the following operators when $|\omega|\geq 3$ (we will come back to $|\omega|=2$ later),
\begin{align}
\langle 1|_1 \tr_{\{3,\dots,|\omega|\}} ww^T| 1\rangle_1 = |1\rangle\langle 1|_2, \\
\langle 1|_1 \tr_{\{3,\dots,|\omega|\}} vv^T| 1\rangle_1 \propto |1\rangle\langle 1|_2.
\end{align}
The transversal gate, represented by a rotation $O$, fixes at least one of $v$ or $w$ ($Ov=v$ or $Ow=w$), so
\begin{align}
|1\rangle\langle 1|_2 & = \langle 1|_1 \tr_{\{3,\dots,|\omega|\}} Oww^TO^T| 1\rangle_1 \nonumber\\
& = O_2\langle 1|_1 \sum_{i=1}^{2^r-1} (O_1\otimes I)|ii\rangle\langle ii|(O_1^T\otimes I) |1\rangle_1 O_2^T \nonumber\\
& = O_2\left[ \sum_{i=1}^{2^r-1} (O_1)^2_{1,i} |i\rangle\langle i|_2\right] O_2^T
\end{align}
or
\begin{align}
|1\rangle\langle 1|_2 & \propto \langle 1|_1 \tr_{\{3,\dots,|\omega|\}} Ovv^TO^T| 1\rangle_1 \nonumber\\
& = O_2\langle 1|_1 \sum_{i=1}^{4^r-1} |\alpha_i|^2(O_1\otimes I)|ii\rangle\langle ii|(O_1^T\otimes I) |1\rangle_1 O_2^T \nonumber\\
& = O_2\left[ \sum_{i=1}^{4^r-1} (O_1)^2_{1,i} |i\rangle\langle i|_2\right] O_2^T.
\end{align}

In the case where $O$ acts on $v$, ``case $v$'', we can conclude that the entire first row of $O_1$ has one nonzero entry, and
the square of this real entry must be $1$. Considering analogous operators, and understanding that $O_j$ is nonsingular, we 
conclude that $O_j$ is a monomial matrix for ``case $v$'', so the corresponding unitary must normalize the Pauli group, i.e.
it must be Clifford.

In the case where $O$ acts on $w$, ``case $w$'', the operator only has rank $1$ if one of $(O_1)_{1,i}$ is nonzero and the rest are 
zero for $i\in [2^r-1]$. However, the equation is only satisfied if the nonzero entry is $\pm 1$ since $O_2$ is an orthogonal matrix.
Therefore, considering analogous operators, $O_j$ has a monomial subblock for ``case $w$'', where $j\in\omega$ and $\omega$ is a minimal 
support, and the south and east subblocks are zero, i.e.
\begin{equation}\label{Eq-Subblock}
O_1 = \left( \begin{matrix} M & 0 \\ 0 & M'\end{matrix}\right),
\end{equation}
where $M$ is a monomial matrix whose nonzero entries are $\pm 1$ and $M'$ is in $SO(4^r-2^r)$.
Therefore, the corresponding unitary matrix must normalize the Pauli Z operators that correspond to the rows and columns of the $M$ matrix.

Therefore, we have the following results. If $\omega$ is a minimal support, $|\omega|\geq 4$, and $A_\omega=3$, then $U_j$ is an $r$ 
qubit Clifford gate for $j\in\omega$. If $A_\omega=1$, and $|\omega|\geq 3$, on the other hand, then up to local Clifford gates $U_j$ 
is an $r$ qubit unitary that normalizes Pauli Z operators but acts arbitrarily on Pauli X operators. In both cases, $U_j$ is a semi-Clifford operation.

The case $A_\omega=3$ and $|\omega|=2$ is a special case. In this case, the minimal subcode is a $[[2,0,2]]$, which we know to be
a Bell pair. The Bell pair is preserved by a continuum of local rotations $U\otimes U^\ast$, so it is an edge case that we
must discard. Since the possible Pauli operators are exhaused on $\omega$, the stabilizer code must be of the form 
$S=S'\otimes\rho_{[[2,0,2]]}$. Therefore, the Bell pair is actually appended to the code and does not improve its ability 
to detect errors on any encoded qubit. If a binary stabilizer code cannot be decomposed as $Q=Q'\otimes [[2,0,2]]$, then 
the code is \textbf{free of Bell pairs}.

The cases $A_\omega=1$ and $|\omega|=1$ or $|\omega|=2$ are special cases as well. In the first case, the qubit at the coordinate $j\in\omega$
is in a product state with the rest of the code. We can discard this case by insisting that $Q\neq Q'\otimes [[1,0,1]]$ is 
\textbf{free of single qubit states}, but this isn't necessary because it is covered by the statements of Theorem 1. In the second case, we do not have enough qubits to ``lock the state to the diagonal" by projecting onto
the first qubit. Therefore, we can only say that 
\begin{align}
O_1\left[\sum_{i=1}^{2^r-1}|i\rangle\langle i|\right]O_1^T & = \tr_2 Ow \nonumber\\
& =\tr_2 w \nonumber\\
& =\sum_{i=1}^{2^r-1}|i\rangle\langle i|, 
\end{align}
i.e. that $U_j$ maps linear combinations of Pauli Z operators to linear combinations of Pauli Z operators. Therefore, in this case, $U_j$ is a generalized semi-Clifford operation.

\subsubsection{Coordinates not covered by minimal subcodes} 

In general, however, a stabilizer code need not be completely described by its minimal elements, i.e. we cannot always find a minimal support containing a coordinate $j\in [n]$. In this section, we place restrictions on the operators $U_j$ in a transversal gate $U = \otimes_{j=1}^n U_j$ when $j$ is not contained in a minimal support.

Suppose we cannot find a minimal support containing coordinate $j$. 
Take the set $S_j:=\{ R\ |\ R\in S(Q), j\in\supp(R)\}$ of stabilizer elements with support on $j$. Since we
assume that the code does not have trivially encoded qubits, $S_j$ is nonempty. Of those elements in $S_j$,
we can single out the set of \textrm{restricted minimal elements} 
$M_j:=\{ R\in S_j\ |\ \nexists R'\in S_j, \supp(R')\subset\supp(R)\}$. Now we can show that if two elements
in $M_j$ have different Paulis at coordinate $j$, then they have different supports. Indeed, suppose there are two
elements $R_a,R_b\in M_j$ that differ on the $j$th coordinate and satisfy $\omega:=\supp(R_a)=\supp(R_b)$. Then
$R_aR_b\in M_j$ and $R_aR_b$, $R_a$, and $R_b$ exhaust the Paulis on the $j$th coordinate. So, up to local Clifford 
operations $R_a=X^{\otimes |\omega|}$ and $R_b=Z^{\otimes |\omega|}$. Since there was no minimal support containing
$j$, there exists some $R\in S\setminus S_j$ such that $\supp(R)\subset\omega$. Furthermore, $R_aR, R_bR, R_aR_bR\in S_j$
because $R\notin S_j$. However, one of these three elements has support strictly contained in $\omega$, contradicting
the definition of $M_j$.

Indeed, suppose the coordinate $j$ is not in any minimal support. Take any $R\in M_j$ and let $\omega=\supp(R)$.
Without loss of generality, suppose $R^{(j)}=Z$. By our previous argument, $\rho_\omega$ contains elements
from $M_j$ that only have Pauli Z at coordinate $j$ and are supported entirely on $\omega$. It also contains elements
from $S$ that have support strictly contained in $\omega$ but have identity at coordinate $j$. In notation,
$\rho_\omega=\sum_i Z_j\otimes R_i + \sum_k I_j\otimes R_k$. Now, we can apply a similar argument to the case
we encountered earlier for $A_\omega=1$ and $|\omega|=2$. The form of the subcode projector is too weak for us
to take a trace over other coordinates, but, like before, we observe that $U_j$ must keep the span of Pauli Z operators invariant, i.e. $U_j$ is a generalized semi-Clifford operation. We have therefore proved Theorem 1 using minimal subcodes.

\subsection{Subcodes associated with single qubits}\label{Sec-OneQubitSubcodesBinary}

In this section we introduce the {\bf single qubit subcodes}, and use these subcodes to prove Theorem 1. This approach provides a more intuitive, accessible proof, as the single qubit subcodes are easier to visualize and understand than the minimal subcodes introduced in Section \ref{Sec-MinimalSubcodesBinary}.

\subsubsection{Single qubit subgroups and subcodes}\label{Sec-SingleQubitSubcodes1}

The {\bf single qubit subcode} associated with a coordinate $i \in [n]$ is the subcode with projector $\rho_{\omega} = \tr_{\bar{\omega}} P_{Q}$, where $\omega = \{i\}$. We denote the projector for this subcode by $\rho_i$. The {\bf single qubit subgroup} $S\langle i \rangle$ associated with $i$ is the set $\{R \in S\ |\ R^{(i)} = I\}$. We define the support of a subgroup $S\langle i \rangle$ to be the set $\cup_{R \in S\langle i \rangle} \supp(R)$. The single qubit subcodes and subgroups have been used by Gross and Van den Nest to study the local unitary and local Clifford equivalence of stabilizer and graph states \cite{Gross}. We will generalize some of their methods to prove Theorem 1.

We begin by reviewing two lemmas by Gross et al.\cite{Gross}. For every subgroup $G$ of $S$, we let $[S:G]$ denote the index of $G$ in $S$. 

{\bf Lemma 3:} Let $S$ be a stabilizer on $n$ qubits, and let $S\langle i \rangle$ denote the single qubit subgroup associated with $i \in [n]$. Then $[S:S\langle i \rangle]= 1, 2$, or $4$ for every $i \in [n]$.

{\bf Lemma 4:} Let $\Pi$ be the smallest subgroup of $S$ containing all the single qubit subgroups $S\langle i \rangle$. We then obtain one of three cases. Either $S = \Pi$, or $[S : \Pi] = 2$, or $[S : \Pi] = 4$. If $\Pi$ has index $4$ in $S$, then the stabilizer code associated with $S$ must be a $[2m, 2m-2, 2]$ code. Note that we can write $\Pi$ as the set $\{R_1R_2\cdots R_n\ | \ R_i \in S\langle i \rangle ,\ i\in [n]\}$.

\subsubsection{Transversal gates on single qubit subcodes}

Following a similar approach to Sec. \ref{Sec-MinSubcode2}, we show that if a coordinate $j \in [n]$ is contained in the support of some single qubit subgroup $S\langle i \rangle$, then the corresponding operator $U_j$ in a transversal gate $U = \otimes_{j=1}^n U_j$ is generalized semi-Clifford. 

We prove the result by induction. If $n=2$, then up to local Clifford equivalence plus permutations of the two qubits the only stabilizer code $Q$ satisfying the requirements of Theorem 1 has the projector
\begin{align}
P_Q &= \frac{1}{2}(I \otimes I + Z \otimes Z).
\end{align}
It is straightforward to verify that the result holds for this code. (See p. 9 in \cite{Bei}. The relevant case is $|\omega|=2$ and $A_{\omega}=1$.)

In the induction step of the proof, let $n \geq 3$ and suppose that the result has been verified for all $n' < n$. Let $Q$ be a stabilizer code on $n$ qubits satisfying the requirements of Theorem 1 and let $U=\otimes_{j=1}^n U_j$ be a transversal gate on $Q$. For every $i \in [n]$, define the set $\omega_i = [n]\setminus \{i\}$. Using Lemma 1, we find that
\begin{align}
U_{\omega_i}\rho_{\omega_i}^{\otimes r}U_{\omega_i}^{\dagger} = \rho_{\omega_i}^{\otimes r},
\end{align}
where $U_{\omega_i}$ is the restriction of $U$ to $\omega_i$ and $\rho_{\omega_i}$ is defined as $\tr_{\bar{\omega_i}} P_Q$. Since $\rho_{\omega_i}$ is the projector for a stabilizer code on $n-1$ qubits, and satisfies the requirements of Theorem 1, we can apply the induction hypothesis to the code corresponding to $\rho_{\omega_i}$ for every $i \in [n]$. This proves that $U_j$ is generalized semi-Clifford for every $j \in [n]$ that is contained in the support of some $S \langle i \rangle$.

\subsubsection{Coordinates not covered by single qubit subcodes}\label{Sec-SingleQubitSubcodes2}

It could be the case that there is a coordinate $j \in [n]$ that is not contained in the support of any $S\langle i \rangle$. However, it is still possible to show that the corresponding operator $U_j$ in a transversal gate $U = \otimes_{j=1}^n U_j$ is generalized semi-Clifford.

Suppose that the coordinate $j$ is not contained in the support of any $S\langle i \rangle$. From the form of $\Pi$ defined in Sec. \ref{Sec-SingleQubitSubcodes1}, we can see that $j \not\in \supp(\Pi)$. It follows that $\Pi$ is strictly contained in $S$. By Lemma 4, $\Pi$ therefore has index 2 or 4 in $S$. If $[S : \Pi] = 4$, then we know that the code $Q$ associated with $S$ is a $[2m, 2m-2, 2]$ code. By Lemma 3 in \cite{Bei}, we find that the transversal gate $U$ on such a code is a local Clifford operation. Thus $U_j$ is a Clifford operation, and therefore generalized semi-Clifford.

If $[S:\Pi] = 2$, then the stabilizer $S$ can be partitioned into two cosets of $\Pi$ as $S = \Pi \cup h\Pi$, where $h \in S\setminus \Pi$. We can see from the definition of $\Pi$ that $h$ has full support. Together with our assumption that $j \not\in \supp(\Pi)$, this implies that for every $R \in S$, we must have $R^{(j)} \in \{I, h^{(j)}\}$. It follows that $[S : S\langle j \rangle] = 2$. We can then partition $S$ into two cosets of $S\langle j \rangle$ as $S = S\langle j \rangle \cup gS\langle j \rangle$, where $g \in S \setminus S\langle j \rangle$.

Defining $\rho\langle j \rangle \equiv \frac{1}{2^n} \sum_{R \in S \langle j \rangle} R$, it follows from the definition of $Q$ that
\begin{align}\label{Eq-1QubitEntireCode}
P_Q = (\underbrace{I\otimes\dots\otimes I}_{n\ \textrm{times}} \ +\ g)\rho\langle j \rangle.
\end{align}
We now compute the projector $\rho_j$ for the single qubit subcode associated with $j$. We find that
\begin{align}
\rho_j &= \sum_{R \in S,\,\supp(R) \subseteq \{j\}} R\nonumber \\
&= I + g^{(j)},
\end{align}
where the second equality follows from the form of $P_Q$ given in Equation (\ref{Eq-1QubitEntireCode}). We can see that $g^{(j)} \in \{X, Y, Z\}$. As we have $U_j \rho_j^{\otimes r} U_j^{\dagger} = \rho_j^{\otimes r}$ by Lemma 1, it follows that $U_j$ maps linear combinations of Pauli $g^{(j)}$ operators to linear combinations of Pauli $g^{(j)}$ operators. Therefore $U_j$ is a generalized semi-Clifford operation. We have thus proved Theorem 1 using single qubit subcodes.

\section{The structure of stabilizer subgroups of stabilizer codes: nonbinary case}\label{Sec-NonBinary}

In many quantum computational problems, the dimension of the computational unit plays an important role. Here, we would like to understand its effect on the set of possible transversal gates. That is, we want to find out, in the qudit settings, whether transversal gates can form a universal set of gates for one of the encoded logical qudits and if not, what operations can be transversal. We will follow a line similar to that in the qubit case but with emphasis on parts that are different and need special notice. First, we study the physical restrictions on transversal gates by analyzing the transformation of stabilizer subcodes under such transversal operations. 

Our main task in this section is to prove the following theorem.

{\bf Theorem 2:} Given a $d$-dimensional $n$-qudit stabilizer code $Q$ free of Bell pairs and trivially encoded qudits, let $U \in I_Q^r$. Then for each $j\in [n]$ either \\
(1) $U_j$ is an $r$ qudit Clifford gate, or\\
(2) $U_j$ keeps a subgroup of the $r$-qudit Pauli group invariant under conjugation, or\\
(3) $U_j$ keeps the span of a subgroup of the $r$-qudit Pauli group invariant under conjugation.

Here by ``Bell pairs" we mean the two-qudit maximally entangled states, which are states locally equivalent to the state $\frac{1}{\sqrt{d}}\sum_{i=0}^{d-1}|ii\rangle$.

For the rest of this section we will work with a $d$-dimensional $n$-qudit stabilizer code $Q$ with corresponding stabilizer $S$ that satisfies the conditions of Theorem 2.

\subsection{Minimal subcodes and beyond}\label{Sec-MinimalSubcodesQudit}

\subsubsection{Minimal subcodes}

In this section we again make use of the technique of minimal subcodes in order to place restrictions on the form of a transversal gate. The generalization of the binary case is mostly straightforward. We continue to use Rains' technique of viewing the projectors onto the codespace as vectors, and the transversal gates as rotations acting on these vectors. However, when $d > 2$ the non-zero entries of the rotation matrices are not necessarily $\pm 1$, but can be any complex number of modulus 1. As a result, the restrictions placed on the form of a transversal gate $U=\otimes_{j=1}^n U_j$ in Theorem 2 differ slightly from those of Theorem 1, stating that $U_j$ preserves the span of a subgroup of the generalized Pauli group under conjugation, rather than a maximal abelian subgroup of the Pauli group. 

As in the binary case, we begin by trying to determine the structure of the projector onto a minimal subcode. Given a minimal support $\omega$, we again use $S_{\omega}$ to denote the subgroup of $S$ generated by the elements of $S$ with support $\omega$. The minimal subcode corresponding to $\omega$ is the code stabilized by $S_{\omega}$. We can list the elements of $S_{\omega}$ as $I, R_1,\dots, R_m$, where 
\begin{align} R_1 &= R_1^{(1)} R_1^{(2)} \dots R_1^{(|\omega|)}\nonumber\\ 
R_2 &= R_2^{(1)} R_2^{(2)} \dots R_2^{(|\omega|)} \nonumber\\ &\ \ \vdots \nonumber\\ 
R_m &= R_m^{(1)} R_m^{(2)} \dots R_m^{(|\omega|)}. \end{align} 

For any Pauli operator $g$, define its order $p$ to be the minimal positive integer that satisfies $g^p = I$. It is easy to see that for each $R_i \in S_{\omega}$, the operators $R_i^{(j)}$ must be of the same order. Otherwise there would exist a certain power $m$ of $R_i$ such that $R_i^m$ had a support strictly contained in $\omega$, contradicting the assumption that $\omega$ is minimal. It can be checked that each Pauli subgroup $\{I, R_1^{(j)}, \dots, R_m^{(j)}\}$ at a particular coordinate $j$ has the same structure, i.e. they have the same multiplication table. This set of subgroups have the same order and their elements correspond. Therefore, up to local Clifford operations, $R_i = (R_i^{(1)})^{\otimes |\omega|}$. Each minimal subcode is then represented by a single-qudit Pauli subgroup $\{I, R_1^{(1)}, \dots, R_m^{(1)}\}$. 
%%%%%%%%%%%%%%%%%%%%%

We can further simplify the form of the minimal subcode. Note that while the operators $R_i$ must commute, the same does not hold for the $R_i^{(1)}$. However, no matter what the commutation factors are for the single-qudit operators, the subcode weight $|\omega|$ is such that they vanish for the $R_i$. Thus we need not concern ourselves with the commutation relations of the Pauli operators  $R_i^{(1)}$ and simply treat them as commutative. In this way, we are dealing with the quotient group ${\mathcal P}_1^d* = {\mathcal P}_1^d/C_{\mathcal P}$, where ${\mathcal P}_1^d$ is the one qudit Pauli group and $C_{\mathcal P} =\{I, q_d I, \dots q_d^{d-1} I\}$ is the center of ${\mathcal P}_1^d$. The group ${\mathcal P}_1^d*$ is then a finite abelian group formed by the direct product of two cyclic-d groups that are generated by X and Z respectively. Its subgroups are of the form $\langle Z^m \rangle$ or $\langle X^{m_1}, Z^{m_2} \rangle$, where $m$, $m_1$ and $m_2$ are factors of $d$. The minimal subcodes are the codes stabilized by these subgroups.

We can now explicitly write out the projectors for minimal subcodes. Denote the number of generators for a subcode by $N_g$. When $N_g = 1$, the $r$ block projector can be written as
\begin{align}
\rho_\omega^{\otimes r} & \propto (I^{|\omega|} + (Z^m)^{|\omega|}+\dots+(Z^{(p-1)m})^{|\omega|} )^{\otimes r}\nonumber\\& = \sum_{i\in \{0\dots p-1\}^r} ((Z^m)^\omega)^{i_1}\otimes \dots \otimes ((Z^m)^\omega)^{i_r} \nonumber\\
& = \sum_{i\in \{0\dots p-1\}^r} Z(i)^{\otimes |\omega|}.\label{quditng1}
\end{align}
This differs from the qubit expression only in that each component of $i$ can take $p$ different values, rather than two ($p$ not necessarily prime). Similarly, the projector $\rho_\omega^{\otimes r}$ when $N_g = 2$ is given by
\begin{align}
\rho_\omega^{\otimes r} & \propto \Big (\sum_{\substack{c\in \{0\dots p_1-1\}\\ d\in \{0\dots p_2-1\}}} ((Z^{m_1})^c (X^{m_2})^d) \Big )^{\otimes r}\nonumber\\
& = \sum_{\substack{a\in \{0\dots p_1-1\}^r\\ b\in \{0\dots p_2-1\}^r}} R^\omega(a_1,b_1)\otimes\dots\otimes R^\omega(a_r,b_r) \nonumber\\
& = \sum_{\substack{a\in \{0\dots p_1-1\}^r\\ b\in \{0\dots p_2-1\}^r}} R(a,b)^{\otimes |\omega|},\label{quditng2}
\end{align}
where $R(a_i,b_i)=(Z^{m_1})^{a_i} (X^{m_2})^{b_i}$ and $R(a,b)=\otimes_{j=1}^r R(a_j,b_j)$.

\subsubsection{Transversal gates on minimal subcodes}

We can now use the techniques of Sec. \ref{Sec-MinSubcode2} to place restrictions on the operators $U_j$ of a transversal gate $U = \otimes_{j=1}^n U_j$ such that $j$ is contained in some minimal support $\omega$. The Pauli group forms a basis for any operator on the $d$ dimensional Hilbert space. Therefore, conjugation of a Pauli operator by tranversal gates can be seen as a unitary transform in the operator space given by
\begin{align} U_jR_sU_j^\dag = \sum_{R_t\in {\mathcal B}_r^d-\{I\}} \alpha_{ts} R_t, \end{align} 
where ${\mathcal B}_r^d$ denotes the basis set (defined in Equation~\ref{oba}) of the $r$-qudit Pauli group. The unitarity of the transformation can be easily proved as in the qubit case. However, unlike the qubit case, $\alpha_{ts}$ is in general a complex number as the Pauli operators $R_t$ are not necessarily Hermitian. Thus we can represent each transversal gate $U_j$ on the code space by a matrix $V_j \in SU(d^{2r}-1)$. We associate the basis elements $\{X^a Z^b \ | \ a,b=0,\dots d-1 \}$ of the generalized Pauli group with the basis vectors $\{|i\rangle \ | \ i=0,\dots d^{2r}-1 \}$. Then the subcode projectors can again be mapped into vectors in $({\mathbb C}^{d^{2r}-1})^{\otimes |\omega|}$.

When $N_g = 1$, we find that
\begin{equation}
\rho_\omega^{\otimes r} - I \mapsto \sum_{i} \underbrace{|ii\dots i\rangle}_{|\omega|\ \textrm{times}} =: w
\end{equation}
The summation is over all vectors $|i\rangle$ that correspond to Pauli matrices $(Z^m)^{i_1}\otimes \dots \otimes (Z^m)^{i_r}$ in Equation~(\ref{quditng1}).

When $N_g = 2$, the mapping takes the same form except that the summation is over all vectors that correspond to Pauli matrices $R(a,b)=\otimes_{i=1}^r (Z^{m_1})^{a_i} (X^{m_2})^{b_i}$ in Equation~(\ref{quditng2}).

Rains' technique still works here to ensure that when $|\omega|\geq 3$, the matrix $V_j$ is either monomial itself or has a monomial subblock as in Equation (\ref{Eq-Subblock}). As mentioned at the beginning of this section, the only difference is that the non-zero entries in the monomial subblock are not necessarily $\pm 1$, but can be any complex number with modulus $1$. Therefore we find that the transversal gate $U_j$ is either Clifford or normalizes a subgroup of the Pauli group.

Now we deal with the case when $|\omega| \leq 2$. As the operators $X^{\otimes |\omega|}$ and $Z^{\otimes |\omega|}$ do not commute for any $d \geq 3$ when $|\omega| \leq 2$, we are only concerned with the case when the Pauli operators at coordinate $j$ are a proper subgroup of all the Pauli operators. When $|\omega| = 2$, we can prove as before that a transversal gate $U_j$ preserves the span of a certain subgroup of the Pauli group under conjugation. When $|\omega| = 1$, if we require that the physical qudit and logical qudit must have the same dimension, we are left only with a trivially encoded qudit--a case that can be discarded.

\subsubsection{Coordinates not covered by minimal subcodes}

Now that we have dealt with the coordinates that are contained in some minimal support, we can go back to see what happens when a $j$th coordinate of the stabilizer code is not covered by any minimal support. As in the qubit case, we remove all the restricted minimal elements $M_j:=\{ R\in S_j\ |\ \nexists R'\in S_j, \supp(R')\subset\supp(R)\}$ from the set $S_j$ of stabilizer elements covering the coordinate $j$. We can again prove, as in the qubit case, that for a fixed support (containing $j$) the Pauli operators at $j$ in the minimal elements form a proper subgroup of the $1$-qudit Pauli group. In this way, we can deduce that $U_j$ must keep the span of a subgroup of Pauli operators invariant under conjugation. We have therefore proved Theorem 2 using minimal subcodes.

\subsection{Subcodes associated with single qudits}

In this section we introduce the {\bf single qudit subcodes}, and use these subcodes to prove Theorem 2. The definitions and results are similar to those of Sec. \ref{Sec-OneQubitSubcodesBinary}, but have been adapted for the case when $d > 2$. The generalization is mostly straightforward, but requires a few adjustments when $d$ is nonprime. The most significant difference lies in the qudit versions of Lemmas 3 and 4, which no longer give specific values for the indices of $S\langle i \rangle$ and $\Pi$ in $S$, but give bounds instead. This slight relaxation still allows us to prove the necessary result.

\subsubsection{Single qudit subgroups and subcodes}\label{Sec-1QuditSubcode1}

The {\bf single qudit subcode} associated with a coordinate $i \in [n]$ is the subcode with projector $\rho_{\omega} = \tr_{\bar{\omega}} P_Q$, where $\omega = \{i\}$. We denote the projector for this subcode by $\rho_i$. The {\bf single qudit subgroup} $S\langle i \rangle$ associated with $i$ is the set $\{R \in S \ | \ R^{(i)} = I\}$. As in the case $d=2$, we define the support of a subgroup $S\langle i \rangle$ to be the set $\cup_{R \in S \langle i \rangle} \supp(R)$.

We will now generalize the two lemmas of Gross et al.\cite{Gross} that we introduced in Sec. \ref{Sec-SingleQubitSubcodes1}.

{\bf Lemma 5:} Let $S$ be a stabilizer on $n$ qudits, and $S\langle i \rangle$ the single qudit subcode associated with $i \in [n]$. Then $[S : S \langle i \rangle] \leq d^2$ for every $i \in [n]$. 

To prove this, note that since $S\langle i \rangle$ is a subgroup of $S$, we can partition $S$ into $N$ cosets of $S\langle i \rangle$ where $N = [S : S \langle i \rangle]$. We can therefore write
\begin{align}
S &= S\langle i \rangle \cup g_1S\langle i \rangle \cup \dots \cup g_{N-1}S\langle i \rangle
\end{align}
for $N-1$ elements $g_1,\dots, g_{N-1} \in S$. Two elements $g_a, g_b \in S$ belong to different cosets of $S\langle i \rangle$ if and only if their $j$th coordinates $g_a^{(j)}$ and $g_b^{(j)}$ differ. Thus, there can be at most $d^2$ cosets of $S\langle i \rangle$, as any basis element $g$ of the generalized Pauli group can be written in the form $Z^{k_1}X^{k_2}$ for $k_1, k_2 \in \{0,1,\dots,d-1\}$. It follows that $[S : S \langle i \rangle] \leq d^2$, and the lemma is proved.

{\bf Lemma 6:} Let $\Pi$ be the smallest subgroup of $S$ containing all the single qudit subgroups $S\langle i \rangle$. Then $[S: \Pi] \leq d^2$. Let $\langle X^{\otimes n}, Z^{\otimes n} \rangle$ be the group generated by $X^{\otimes n}$ and $Z^{\otimes n}$. If $[S : \Pi] = d^2$, then the stabilizer $S$ can be written up to local Clifford operations as $\langle X^{\otimes n}, Z^{\otimes n} \rangle$, where $X$ and $Z$ are the generators of the generalized $r$-qudit Pauli group.

To prove the first part of the lemma, we use the fact that $|S| = |G|[S : G]$ for any subgroup $G$ of $S$. As every single qudit subgroup $S\langle i \rangle$ is contained in $\Pi$, it follows that $|S\langle i \rangle| \leq |\Pi|$ for every $i \in [n]$. Thus, we find that $[S : \Pi] \leq [S : S\langle i \rangle] \leq d^2$. 

To prove the second part of the lemma, assume that $[S : \Pi] = d^2$. As in the case $d=2$, we can write $\Pi$ as the set $\{R_1R_2\dots R_n \ | \ R_i \in S \langle i \rangle, i \in [n]\}$. We can partition $S$ into $d^2$ cosets of $\Pi$:
\begin{align}
S &= \Pi \cup g_1\Pi \cup \dots \cup g_{d^2-1}\Pi,
\end{align}
for $d^2-1$ elements $g_1,\dots, g_{d^2-1} \in S$. It follows from the definition of $\Pi$ that every $g_k$ must have full support. The $g_k$ must also differ pairwise on every qudit. To see this, assume that $g_{k_1}^{(m)} = g_{k_2}^{(m)}$ for some pair $k_1, k_2$, and let $g^{(m)} := g_{k_1}^{(m)}$. Let $p$ denote the order of $g^{(m)}$. Then since $I^{\otimes n} = (g^{(m)})^p$, it follows that $g_{k_1}^{p-1}g_{k_2} \in \Pi$. The element $g_{k_1}g_{k_1}^{p-1}g_{k_2}$ belongs to the coset $g_{k_1}\Pi$. But the element $g_{k_1}g_{k_1}^{p-1}g_{k_2}$ also belongs to the coset $g_{k_2}\Pi$. Thus we have $g_{k_1}\Pi = g_{k_2}\Pi$, and therefore $k_1 = k_2$. It follows that the $g_k$ differ pairwise on every qudit.

We now show that the only element in $\Pi$ is $I^{\otimes n}$, which immediately implies that $S = \{I^{\otimes n}, g_1, \dots, g_{d^2-1}\}$. Assume that there is an element $f \in \Pi$ such that $f^{(m)} \neq I$ for some $m \in [n]$. Then $f^{(m)} = g_k^{(m)}$ for some $k \in \{1,\dots, d^2-1\}$. Let $f^{(m)}$ have order $p$. Then we find that $f^{p-1}g_k \in \Pi$. Let $f$ have order $p'$. Then $f^{p' - (p-1)}(f^{p-1}g_k) = g_k \in \Pi$. But this is a contradiction, as $g_k$ is an element of $g_k\Pi$, which is a coset of $\Pi$ disjoint from $\Pi$. It follows that $f = I^{\otimes n}$, and therefore $\Pi = \{I^{\otimes n}\}$ and $S = \{I^{\otimes n}, g_1, \dots, g_{d^2-1}\}$. As the elements $g_k$ have full support and differ pairwise on every qudit, we find that $S$ can be written up to local Clifford operations as $\langle X^{\otimes n}, Z^{\otimes n} \rangle$, where $X$ and $Z$ are the generators of the generalized $r$-qudit Pauli group. The lemma is proved.

\subsubsection{Transversal gates on single qudit subcodes}

In this section we show that if a coordinate $j \in [n]$ is contained in the support of some single qudit subgroup $S\langle i \rangle$, then the corresponding operator $U_j$ in a transversal gate $U = \otimes_{j=1}^n U_j$ preserves the span of a subgroup of the generalized $r$-qudit Pauli group under conjugation.

We prove the result by induction. If $n=2$, let $S$ be the stabilizer of a code $Q$ satisfying the conditions of Theorem 2. Every element $R \in S$ must be of the form $R = R^{(1)} \otimes R^{(2)}$, where $R^{(1)}$ and $R^{(2)}$ have the same order. If they were not of the same order, then $S$ would contain an element of weight 1, contradicting the assumptions on $Q$. As $Q$ is free of Bell pairs, the set $\{R^{(1)} \ | \ R \in S\}$ does not form the entire Pauli group. We can then follow the proof for weight 2 subcodes in Sec. \ref{Sec-MinimalSubcodesQudit} to conclude that $U_j$ preserves the span of a subgroup of the generalized Pauli group for $j=1,2$. Thus the theorem holds in the case $n=2$.

The induction step of the proof is identical to the case when $d=2$. Therefore, if a coordinate $j \in [n]$ is contained in the support of some $S\langle i \rangle$, then the corresponding operator $U_j$ of a transversal gate $U = \otimes_{j=1}^n U_j$ preserves the span of a subgroup of the generalized Pauli group under conjugation.

\subsubsection{Coordinates not covered by single qudit subcodes}

Following the approach of Sec. \ref{Sec-SingleQubitSubcodes2}, we consider the case when a coordinate $j \in [n]$ is not contained in the support of any $S \langle i \rangle$, and show that the corresponding operator $U_j$ in a transversal gate $U = \otimes_{j=1}^n U_j$ preserves the span of a subgroup of the generalized $r$-qudit Pauli group under conjugation.

Suppose that the coordinate $j$ is not contained in the support of any $S \langle i \rangle$. From the form of $\Pi$, we can see that $j \not\in \supp(\Pi)$. It follows that $\Pi$ is strictly contained in $S$, so by Lemma 6 we know that $2<[S:\Pi]\leq d^2$. If $[S : \Pi] = d^2$, then we know from Sec. \ref{Sec-1QuditSubcode1} that $S = \langle X^{\otimes n}, Z^{\otimes n} \rangle$ up to local Clifford operations. This corresponds to one of the cases outlined in Sec. \ref{Sec-MinimalSubcodesQudit} (the case $N_g = 2$). We can therefore use the methods in this section to show that $U_j$ keeps the span of a subgroup of the generalized Pauli group invariant under conjugation.

If $[S : \Pi] < d^2$, then $S$ can be partitioned into $N = [S : \Pi]$ cosets of $\Pi$ as shown below.
\begin{align}
S = \Pi \cup h_1\Pi \cup \dots \cup h_{N-1}\Pi
\end{align} 
All the elements $h_k \in S\setminus \Pi$. We can see from the definition of $\Pi$ that every $h_k$ has full support. Together with our assumption that $j \not\in \supp(\Pi)$, this implies that for every $R \in S$, we must have $R^{(j)} \in \{I, h_1^{(j)}, \dots, h_{N-1}^{(j)}\}$. It follows that $[S : S\langle j \rangle] = N$ for some $2 \leq N \leq d^2 - 1$. 

We can then partition $S$ into $N$ cosets of $S \langle j \rangle$ as
\begin{align}
S = S\langle j \rangle \cup g_1S\langle j \rangle \cup \dots \cup g_{N-1}S\langle j \rangle,
\end{align}
where each element $g_k \in S\setminus S\langle j \rangle$.  

Defining $\rho\langle j \rangle \equiv \frac{1}{2^{nd}} \sum_{R \in S\langle j \rangle} R$, it follows from the definition of $Q$ that
\begin{align}\label{Eq-QuditEntireCode}
P_Q = (I^{\otimes n} + g_1 + \dots + g_{N-1})\rho \langle j \rangle.
\end{align}
We now compute the projector $\rho_j$ for the single qubit subcode associated with $j$. We find that
\begin{align}
\rho_j &= \sum_{R \in S,\,\supp(R) \subseteq \{j \}} R\nonumber \\
&= I^{\otimes n} + g^{(j)}_1 + \dots + g^{(j)}_{N-1},
\end{align}
where the second equality follows from the form of $P_Q$ given in Equation (\ref{Eq-QuditEntireCode}). As we have $U_j\rho_j^{\otimes r}U_j^{\dagger} = \rho_j^{\otimes r}$ by Lemma 1, it follows that $U_j$ preserves the span of a subgroup of the generalized Pauli group under conjugation. The subgroup in question is generated by the set $\{ g_1^{(j)}(i), \dots, g_{N-1}^{(j)}(i) \ | \ i \in \{0,1\}^r \}$, where as before, we use $g(i)$ to denote a Pauli $g$ operator. We have therefore proved Theorem 2 using single qudit subcodes.

\section{Transversal gates on $r$ encoded blocks}\label{Sec-Logical}

In this section we prove that the transversal gates on a stabilizer code $Q$ cannot form an encoded quantum computationally universal set for even one of the encoded qudits. Our proof proceeds by contradiction: we begin by assuming that universality can be achieved on a particular encoded qudit. In particular, we assume that the Hadamard and Phase gates can be implemented transversally. Next, we use these gates to construct logical Pauli operations on the encoded qudit, and show that these operations have minimal support $\omega$. The restrictions on the form of transversal gates given by Theorems 1 and 2 ensure that we can use these logical Paulis and the Hadamard or Phase gate to construct another logical Pauli operator with support strictly contained in $\omega$. This contradicts the fact that $\omega$ is a minimal support. As the only assumption we have made is that the set of transversal gates is universal for a particular encoded qudit, we conclude that this assumption is false and no such set of transversal gates exists.

\subsection{Binary case}

We first consider the case when $d=2$. Recall what we found in Sec. \ref{Sec-Binary}: 
Let $U$ be an element of $I_{Q}^r$ free of Bell pairs and trivially encoded qubits.
Then for each $j\in [n]$, $U_j$ is an $r$-qubit generalized semi-Clifford operation. To be more precise, there are three possibilities: (i) $U_j$ is a Clifford operation if all three Pauli operations $\{X_j,Y_j,Z_j\}$ appear in some minimal subcodes containing the coordinate $j$; (ii) $U_j$ is a semi-Clifford operation if only one of the three Pauli operations $\{X_j,Y_j,Z_j\}$ appears in all the minimal subcodes containing the coordinate $j$, and all those minimal subcodes are of weights greater than $2$; (iii) $U_j$ is a generalized semi-Clifford operation if (a) only one of the three Pauli operations $\{X_j,Y_j,Z_j\}$ appears in all the minimal codes containing the coordinate $j$, and all those minimal subcodes are weight $2$, or (b) The $j$th qubit is not covered by any minimal subcodes. 

With such a restriction on the possible form of $U_j$, we need to understand how this restriction is related to the restrictions of the allowable transversal logical operations on the code $Q$. We have not yet introduced a basis for the logical operators of $Q$, so the discussion to this point applies to both subsystem and subspace codes. However, as we proceed, we should take care when working with logical operators so that our arguments continue to hold for subsystem codes.

We have observed that many transversal gates are Clifford gates, so these gates map logical operators in the 
Pauli group back into the Pauli group. However, it is possible that some transversal gates do not map Paulis to Paulis. 
At first this may seem surprising because we are so familiar with doubly-even dual-containing CSS codes such as the 
$[[7,1,3]]$ Steane code \cite{Steane} and the $[[23,1,7]]$ Golay code \cite{Reichardt}. Codes such as these have 
transversal Phase $\bar{P}$ and 
Hadamard $\bar{H}$ gates implemented bitwise (i.e. by applying said gate or its conjugate to each bit of the code). 
Therefore, all of their minimal subcodes have $A_\omega=3$, and all of their transversal gates are Clifford (they are 
a subset of the GF(4)-linear codes). These codes were designed this way -- they have transversal encoded CNOT, H, and P, 
so we can do any logical Clifford operation transversally. However, there are many examples where codes exhibit 
non-Clifford transversal gates. The $[[9,1,3]]$ Shor code \cite{ShorCode} has a basis
\begin{equation}
|0/1\rangle \propto (|000\rangle+|111\rangle)^{\otimes 3} \pm (|000\rangle -|111\rangle)^{\otimes 3},
\end{equation}
so any gate of the form $e^{i\theta Z_1}e^{-i\theta Z_2}$ preserves the code space and acts as the encoded identity gate.
In other words, this gate is in the \textbf{generalized stabilizer}, which is the set of all unitary gates that
fix the code space \cite{Guys}. Furthermore, the gate is an element of $I_Q^{r}$, the {\bf stabilizer subgroup} defined in Sec. IIB., 
the set of all transversal gates fixing the code space. The $[[15,1,3]]$ CSS code constructed from the punctured 
Reed-Muller code $RM^\ast(1,4)$ and its even subcode has a transversal $\pi/8$-gate $T$ \cite{KnillInject}. This gate is 
implemented by bitwise application of $T^\dag$ and maps the logical Pauli X operator $\bar{X}=X^{\otimes 15}$ to 
$(\frac{X+Y}{\sqrt{2}})^{\otimes 15}$. The image differs from $(\bar{X}-\bar{Y})/\sqrt{2}$ by an element of the local 
identity.

In our proof, we will apply transversal gates that may not take Paulis to 
Paulis, even if the transversal gate (approximates) a logical Clifford gate. These gates may take us outside of the 
stabilizer formalism and force us to deal with rather foreign objects such as the stabilizer subgroup $I_Q$.
Fortunately, we will see that it is possible to remain within the powerful stabilizer formalism.

Partition the logical Pauli operations into two sets, the set of operations on \textbf{protected qubits} and the set of operations on \textbf{gauge qubits}, as defined in Sec. \ref{Sec-Prelim2}. We wish to compute on the protected qubits up to operators on the gauge qubits. We therefore assume that any single qudit logical gate on a protected logical qubit $p$ can be approximated to any accuracy using only transversal gates. 

Let $\alpha$ be a \textbf{minimum weight element} of the union of cosets 
$\bar{X}_p^{(1)}S\cup\bar{Z}_p^{(1)}S\cup\bar{Y}_pS$, where ``$(1)$'' denotes the 
first block. Let $\omega\equiv\supp(\alpha)$. The notation $\bar{X}_p^{(1)}S$ indicates the set of representatives of 
$\bar{X}_p^{(1)}$ in the Pauli group. We are also free to apply any operator to the gauge qubits in the first block 
when choosing our representation $\alpha$, but we know that in doing so, we cannot construct a logical operator on a 
protected qubit that has weight less than $|\omega|$, so this freedom can be safely ignored. Likewise, it does not matter how we represent the identity on blocks, since we must transform all representations correctly. We 
choose to represent it by tensor products of identity operators. 

By our assumption, $\bar{H}_p^{(1)}$ is transversal. On the other blocks, we would like to apply a logical identity gate 
on the protected logical qubits, but again we are free to apply any logical operation to the gauge qubits. Applying 
this gate to $\alpha\otimes I$, we get $\beta''\equiv\bar{H}_p(\alpha\otimes I)\bar{H}_p^\dag$. The operator $\beta''$ 
must represent $\bar{Z}_p^{(1)}$ up to elements of the transversal identity and gauge operators. Expanding $\beta''$ in 
the basis of Pauli operators gives
\begin{align}
\beta'' &= \sum_{R\in {\mathcal P}_{n}^{\otimes r}} \alpha_R R\nonumber\\ &= \sum_{R\in C(S)^{\otimes r}} \alpha_R R + \sum_{R\in {\mathcal P}_{n}^{\otimes r}-C(S)^{\otimes r}} \alpha_R R.
\end{align}
Here $C(S)$ is the centralizer of $S$. The operators not in $C(S)^{\otimes r}$ map the code space to an orthogonal 
subspace, so there must be terms in the expansion that are in $C(S)^{\otimes r}$. Let $\beta'\equiv P_Q\beta'' P_Q$. 
All the terms of the operator $\beta'$ are in $C(S)^{\otimes r}$.  Considering how $\beta'$ acts on a basis of 
$Q^{\otimes r}$, we can neglect terms in $S^{\otimes r}$ because they act as the identity. Therefore, there must be 
an element of $C(S)^{\otimes r}$ that represents $\bar{Z}_p^{(1)}$ and enacts an arbitrary logical Pauli operation 
on the gauge qubits. The transversal gate cannot cause $\beta''$ to have support on the first block that strictly 
contains $\omega$, nor can it have support strictly contained in $\omega$, since $|\omega|$ is minimal. Furthermore, 
$I\in C(S)$ so we can ignore blocks other than the first by finding an operator $\beta$ that represents $\bar{Z}_p^{(1)}$ 
and enacts an arbitrary logical Pauli operation on the gauge qubits in the first block. We also have
$\omega=\supp(\alpha)=\supp(\beta)$. Since there must be some overlap between the operator $\bar{H}_p^{(1)}$
and the centralizer $C(S)$, this line of reasoning holds even if $\bar{H}_p^{(1)}$ is $\epsilon$-close to a transversal 
gate but is not exactly implemented by a transversal gate. Repeating the argument for $\bar{P}_p^{(1)}$, we obtain an 
operator $\gamma$ with support $\omega$ that represents $\bar{Y}_p^{(1)}$ up to logical Paulis on the gauge qubits.

Now we can derive the contradiction. Since we have assumed that the transversal gates are a universal set for some protected qubit $p$, there must be some coordinate $j\in\omega$ such that $U_j$ is not Clifford in the tensor product decomposition of $\bar{H}_p^{(1)}$ or $\bar{P}_p^{(1)}$. Otherwise, we could not apply any non-Clifford logical gates to the encoded qubit $p$. By the restrictions we derived in Sec. \ref{Sec-Binary}, $U_j$ must be semi-Clifford or generalized semi-Clifford. If $U_j$ is semi-Clifford, it must fix one of the Pauli operators at coordinate $j$ in the first block, or it must map one of the Pauli operators to the identity. For example, we could have $U_jZ_1U_j^\dag=\pm Z_1$ or $U_jZ_1U_j=I_1$. Therefore, one of
the images or a product of one of the images of $\alpha$, $\beta$, or $\gamma$ under $\bar{H}_p^{(1)}$ and another 
logical Pauli operator $\alpha$, $\beta$, or $\gamma$ will have support strictly contained in $\omega$, but will also
represent a logical Pauli on the protected qubit. This is impossible because $\alpha$, $\beta$, and $\gamma$ 
already have minimum weight. Thus $U_j$ cannot be semi-Clifford.

Now we can complete our proof by showing that the universality of transversal gates is contradictory to the last possibility, i.e. $U_j$ is generalized semi-Clifford. We can assume without loss of generality that $U_j$ keeps the span of Pauli $Z$ operators invariant. As shown above, there exist three Pauli operators $\alpha$, $\beta$, $\gamma$ $\in C(S)$ which have the same minimum support $\omega$ and are representatives of $\bar{X}_p^{(1)}$, $\bar{Y}_p^{(1)}$, $\bar{Z}_p^{(1)}$ repectively. Because they are of the same minimum support, it can be shown that $\alpha$, $\beta$, $\gamma$ are locally Clifford equivalent to $X^{\otimes |\omega|}$, $Y^{\otimes|\omega|}$, $Z^{\otimes|\omega|}$. Without loss of generality, assume that $\gamma \sim Z^{\otimes|\omega|}$. By our assumption on the universality of transversal gates, both $\bar{H}_p^{(1)}$ and $\bar{P}_p^{(1)}$ are transversal and preserve the span of Pauli $Z$ operators. Thus we have $\alpha'$ and $\beta'$ representing $X^{\otimes|\omega|}$ and $Y^{\otimes|\omega|}$ and of the diagonal form on the $j$th coordinate. Following our previous reasoning we can show that $P_Q \alpha' P_Q$, $P_Q \beta' P_Q$, and $P_Q \gamma P_Q$ also represent $\bar{X}_p^{(1)}$, $\bar{Y}_p^{(1)}$, $\bar{Z}_p^{(1)}$, and that one of them must have support strictly contained in $\omega$. This contradicts the minimality of $\omega$. The only assumption we have made is that the set of transversal gates is universal for the arbitrarily chosen protected qubit $p$, so this assumption must be false.

\subsection{Nonbinary case}

The restrictions on the form of transversal gates that we obtained in Sec. \ref{Sec-NonBinary} limit the range of possible logical operations that we can apply to any stabilizer code. We now prove that, in the general qudit case, universal logical computation is still not possible using only transversal gates on subspace or subsystem stabilizer codes. In the binary case, we proved our result by using the fact that the restricted form of the group of transversal gates prevents them from approximating some logical Clifford operations unless the transversal gates are all contained in the Clifford group. This is no longer the case when $d$ is nonprime, so the generalization of our proof to the qudit case is not trivial. But this does not affect our final conclusion, as shown below. 

The minimum weight element in $C(S)\setminus S$ representing logical Pauli operations $\{\bar{G}_p\}$ on a particular encoded qudit $p$ will help us again in the proof. Suppose that such an element has support $\xi$ and is of order $q$. (For subsystem codes, we can apply any operation to the gauge qudits but this freedom does not affect our choice of minimum weight element, as shown in the qubit section.) We can easily see that on each coordinate within $\xi$ this element has a Pauli operator of order $q$ while all the operators on coordinates outside of $\xi$ are the identity. Up to a local Clifford operation we can write this element as $(X^m)^{\otimes |\xi|}$, where $m\cdot q=d$. Choose this element to represent the logical gate $\bar{X}^m_p$. 

We can show that the generating set $\{\bar{X}_p, \bar{Z}_p\}$ of the logical Pauli group $\{\bar{G}_p\}$ can also be represented on support $\xi$. Our discussion here is up to the same local Clifford operation of $\bar{X}^m_p$. First note that $X^{\otimes |\xi|}$ is also in $C(S)\setminus S$, as otherwise $(X^m)^{\otimes |\xi|}$ cannot be a logical operation. We can therefore assign $X^{\otimes |\xi|}$ to represent $\bar{X}_p$. Under our assumption, all logical Clifford operations are transversal. Thus $\bar{Z}_p$ is represented by $Z^{\otimes |\xi|}$ up to local Clifford operations. Now a whole set of logical Pauli operators $S_{\xi} = \langle X^{\otimes |\xi|}, Z^{\otimes |\xi|}\rangle$ can be generated on support $\xi$.  Each logical Pauli operation $\bar{g}$ is represented by $g^{\otimes |\xi|}$ up to a local Clifford operation.

With such a basis, first we reason that non-Clifford transversal gates are always needed to perform non-Clifford logical operations. Remember that the restrictions we have on non-Clifford transversal gates are: (i) they preserve a subgroup of the physical Pauli operators, or (ii) they preserve the span of a subgroup of the physical Pauli operators. As case (i) is included in case (ii), it is sufficient to show that the second restriction does not allow universal logical operations on any encoded qudit. 

In the qubit case, conditions (i) and (ii) imply that non-Clifford transversal gates are either a semi-Clifford operation or a generalized semi-Clifford operation as any abelian subgroup of the qubit Pauli group is maximal. As previously stated, we proved the main result in the previous section from the fact that (generalized) semi-Clifford operations cannot perform Clifford operations. However, in cases when the dimension $d$ is not prime, Clifford operations might not be excluded by conditions (i) or (ii). For example, when $d=4$, any Clifford operation preserves the subgroup generated by $X^2, Z^2$. In these cases, our previous proof technique will not work--we need to find a new contradiction that is independent of the dimension.

Denote the subgroup whose span is preserved by transversal gates on coordinate $j$ by $P_s$. Choose a logical operation $\bar{A}_p$ that maps operators within the span of $\bar{P}_{s(p)}$ to the outside. The operator $\bar{A}_p$ may contain any operation on the gauge qudits. It is transversal according to our assumption and takes the form $A_1 \dots A_{|\xi|}$. We can write

\begin{equation} \bar{A}_p \bar{\alpha} \bar{A}_p^{\dagger} = \bar{\beta} \end{equation} 
where $\bar{\alpha}$ is some element of $\bar{P}_{s(p)}$ while $\bar{\beta}$ lies outside the span of $\bar{P}_{s(p)}$. Expanding $\bar{\beta}$ in Pauli basis gives

\begin{equation} \bar{\beta} = \bar{\beta_1} + \bar{\beta_2} + \dots + \bar{\beta_1'} + \bar{\beta_2'} + \dots, \end{equation} 
where the $\bar{\beta}_i$'s are in $\bar{P}_{s(p)}$ and the $\bar{\beta}'_i$ are not. With the established correspondence between $\bar{g}_p$ and $g^{\otimes |\xi|}$, we can write (up to local Clifford operations and gauge operations) 

\begin{align} &(A_1 \dots A_{|\xi|}) (\alpha)^{\otimes |\xi|} (A_1^{\dagger} \dots A_{|\xi|}^{\dagger}) \nonumber\\ 
= &(\beta_1)^{\otimes |\xi|} + (\beta_2)^{\otimes |\xi|} + \dots + (\beta_1')^{\otimes |\xi|} + (\beta_2')^{\otimes |\xi|} + \dots
\end{align} 

On the $j$th coordinate accordingly we have 

\begin{equation} \beta = A_j \alpha A_j^{\dagger} \end{equation} 
When expanded in the Pauli basis, $\beta$ must have a component outside of $P_s$, as otherwise there cannot be $(\beta'_i)^{\otimes |\xi|}$'s in the expansion of $\bar{\beta}$. However this contradicts the requirement that $A_j$ keeps the span of $P_s$ invariant. Thus, the assumption that transversal gates are universal must be false in the general qudit case.

\section{Conclusion and open questions}\label{Sec-Conclusion}

In this paper we generalize our work in \cite{Bei} to show that for subsystem stabilizer codes in $d$ dimensional Hilbert space, a universal set of transversal gates cannot exist for even one encoded qudit, for any dimension $d$, prime or nonprime. 

The most natural and important route of investigation at this point is determining to what extent we must continue to strengthen ``transversality'' before we achieve universality. For example, the case where we can permute the bits in addition to carrying out transversal gates is still open. This particular case is of great interest, as it could allow us to simplify the architecture of fault-tolerant quantum computers. However, preliminary investigations suggest that these conditions are still insufficient to achieve universality. Here, we prove that this case does not give universality for a single block binary stabilizer code.

\begin{figure}[htb!]
\centering
\includegraphics[width=2in]{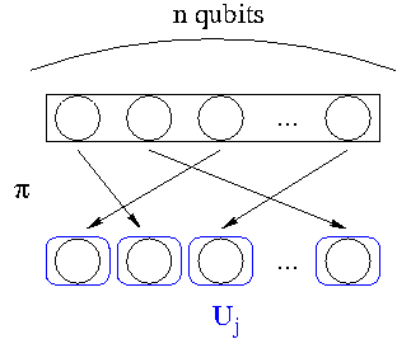}
\caption{Illustration of a code automorphism on $1$ block of $n$ qubits. The block is represented by
a collection of circles (qubits), grouped into a box. The block undergoes a coordinate permutation $\pi$
followed by a local unitary gate $U$ whose unitaries $U_j$ act on qubits in the [blue] boxes with rounded 
edges.}
\label{fig:automorphism}
\end{figure}

An $r$ block \textbf{code automorphism} is a gate of the form $UP_\pi$ that commutes with 
$P_Q^{\otimes r}$, where $U$ is a local unitary gate on all $nr$ qubits and $\pi$ is a coordinate 
permutation of all $nr$ coordinates \cite{RainsAut}. This is illustrated for $r=1$ in FIG.~\ref{fig:automorphism}.
Code automorphisms form a group denoted by $\aut(Q^{\otimes r})$. 

We will show that the code automorphisms on $r$ encoded blocks do not form a universal set for even one encoded qubit.
Since we can regard $Q^{\otimes r}$ as just another code, it is enough to
demonstrate the result for the case of one encoded block, when $r=1$. We will rely on the discussion in Sec. \ref{Sec-Logical}.

As before, let $\alpha$ be a minimum weight element of $C(S)\setminus S$ representing $\bar{X}_p$ without
loss of generality. Let $\omega\equiv\supp(\alpha)$. Consider the single qubit gate $A$ defined by
\begin{align}
X & \mapsto \frac{1}{\sqrt{3}}(X+Y+Z), \nonumber\\
Z & \mapsto Z',
\end{align}
where $\{AXA^\dag,AZA^\dag\}=0$. 

As before, assume that $\bar{A}_p$ is implemented to accuracy $\epsilon$  by some gate $UP_\pi\in\aut(Q)$.
Then $\eta\equiv\bar{A}_p\alpha\bar{A}_p^\dag$ is an element of $\frac{1}{\sqrt{3}}(\bar{X}+\bar{Y}+\bar{Z})\mathcal{I}$, 
where $\mathcal I$ is the generalized stabilizer (not the local identity, since the permutation is not local). 
Expanding $P_Q\eta P_Q$ in the Pauli basis, we again see that there must be representatives $\alpha'$, $\beta'$, 
and $\gamma'$ of $\bar{X}_p$, $\bar{Z}_p$, and $\bar{Y}_p$ in the centralizer $C(S)$ that all have support 
$\omega'$ such that $|\omega'|=|\omega|$. As in Sec. \ref{Sec-Logical}, this is partly because $\alpha$ has minimum weight. The new feature is that $\alpha'$, $\beta'$, and $\gamma'$ must have the same support even though we have applied a permutation.

Now, $U$ must be a local equivalence between $Q'\equiv P_\pi Q$ and $Q$. Thus each $U_j$ is, as before, either a single 
qubit Clifford gate or a gate of the form $L_1e^{i\theta Z}L_2$, where $L_1,L_2$ are single qubit Cliffords. If every $U_j$ is Clifford, then we are done. Otherwise, one or more gates are of the second form. In this case we can assume that $j$ is 
in $\omega''\equiv P_\pi \omega'$ (otherwise $\bar{A}_p$ is Clifford). Let $\delta'$ be another name 
for the Pauli operator in $\{\alpha',\beta',\gamma'\}$ whose $j$th coordinate does not change when we apply $\bar{A}_p$. 
Then $\eta'\equiv\bar{A}_p\delta'\bar{A}_p^\dag$ yields three new Pauli operators with support $\omega''$. At least two of these Pauli operators must have the same Pauli at coordinate $j$, so their product's support is strictly contained in $\omega''$. This contradicts the minimality of $\omega''$. Therefore the gate $\bar{A}_p$ cannot be implemented arbitrarily well by a product of gates in 
$\aut(Q)$. We conclude that $\aut(Q)$ cannot be a universal set.

This result suggests that allowing permutations in addition to transversal gates will still be insufficient to achieve universality. However, our proof cannot be directly generalized to the multiblock case and the qudit case. In the former case, we might allow different permutations on different blocks. In the latter case, it is not clear whether or not we could find a gate similar to the gate $A$ used in our proof that maps

\begin{align}
X & \mapsto \frac{1}{\sqrt{N_x}}\sum\limits_{g\in B}\alpha_g g, \nonumber\\
Z & \mapsto Z',
\end{align}
where $N_x$ is some normalization constant and $\alpha_g\neq 0$ for all $g$ in the generalized Pauli group except the identity.

Several other generalizations could also be considered. For example, we could allow different blocks to be encoded using different codes. We may even be able to use different codes for the input and output. It is clear that allowing the use of measurement immediately gives universality by using teleportation, so we should explore the possibility of using protocols weaker than this to achieve universality on stabilizer codes.

%%%%%%%%%%%%%%%%%%%%%%%%%%%%%%%%%%%%%%%%%%%%%%%%%%%%%%%%%%%%%%%%%%%%%%%%%%%%%

\end{document}